\documentclass[12pt,preprint]{aastex}

\newcommand{\myemail}{lic@nju.edu.cn}

\shorttitle{WTD of SEP events}
\shortauthors{Li et al.}

\begin{document}


\title{Waiting time distribution of solar energetic particle events modeled with a non-stationary Poisson process}


\author{C. Li\altaffilmark{1,2}, S. J. Zhong\altaffilmark{3}, L. Wang\altaffilmark{4}, W. Su\altaffilmark{1,2}, and C. Fang\altaffilmark{1,2}}
\altaffiltext{1}{School of Astronomy and Space Science, Nanjing University, Nanjing 210093, China. \myemail}
\altaffiltext{2}{Key Laboratory for Modern Astronomy and Astrophysics (Nanjing University), Ministry of Education, Nanjing 210093, China}
\altaffiltext{3}{Department of Mathematics, Southeast University, Nanjing 210096, China}
\altaffiltext{4}{Institute of Space Physics and Applied Technology, Peking University, Beijing 100871, China}


\begin{abstract}
We present a study of the waiting time distributions (WTDs) of solar energetic particle (SEP) events observed with the spacecraft $WIND$ and $GOES$. Both the WTDs of solar electron events (SEEs) and solar proton events (SPEs) display a power-law tail $\sim \Delta t^{-\gamma}$. The SEEs display a broken power-law WTD. The power-law index is $\gamma_{1} =$ 0.99 for the short waiting times ($<$70 hours) and $\gamma_{2} =$ 1.92 for large waiting times ($>$100 hours). The break of the WTD of SEEs is probably due to the modulation of the corotating interaction regions (CIRs). The power-law index $\gamma \sim$ 1.82 is derived for the WTD of SPEs that is consistent with the WTD of type II radio bursts, indicating a close relationship between the shock wave and the production of energetic protons. The WTDs of SEP events can be modeled with a non-stationary Poisson process which was proposed to understand the waiting time statistics of solar flares (Wheatland 2000; Aschwanden $\&$ McTiernan 2010). We generalize the method and find that, if the SEP event rate $\lambda = 1/\Delta t$ varies as the time distribution of event rate $f(\lambda) = A \lambda^{-\alpha}exp(-\beta \lambda)$, the time-dependent Poisson distribution can produce a power-law tail WTD $\sim \Delta t^{\alpha - 3}$, where $0 \leq \alpha < 2$.
\end{abstract}


\section{Introduction}

Solar energetic particles (SEPs) are mainly accelerated during the processes of solar eruptions, namely flares and coronal mass ejections (CMEs), with a part of them may arise from other solar activities, e.g., jets or magnetic reconfiguration in high coronal sites (Klein et al. 2001; Pick et al. 2006). The relationship between SEPs and solar eruptions is still not quite understood, even though a large quantity of efforts had been devoted to the studies of the origins of SEPs (for some recent studies refer to Aschwanden 2012; Golpalswamy et al. 2012; Li et al. 2012; Li et al. 2013; Miroshnichenko et al. 2013; Reames 2013).

The times between events, waiting times $\Delta t$, can provide critical information on how individual event works, for instance, whether a event is independent or connected/triggered by the other one. The waiting time distributions (WTDs) are intensively applied in geophysics and astrophysics (Sotolongo-Costa et al. 2000; Lepreti et al. 2004; Wang $\&$ Dai 2013). In solar physics, WTDs had been studied in last two decades for flares, CMEs, and radio bursts, etc, and a consensus has been reached that the WTDs display power-law-tail profiles (Pearce et al. 1993; Wheatland et al. 1998; Wheatland 2003; Eastwood et al. 2010).

The avalanche model of solar eruptions is described as a system of self-organized criticality (SOC) that predicts WTD is a simple exponential profile consistent with a Poisson process (Bak et al. 1988; Lu $\&$ Hamilton 1991; Aschwanden 2012). Different interpretations were suggested to understand the fact of the power-law WTD (Boffetta et al. 1999; Lepreti et al. 2001; Greco et al. 2009), and one of them is the time-dependent or non-stationary Poisson process (Wheatland 2000; Aschwanden $\&$ McTiernan 2010). If the rate of solar eruptions varies with time, the superposition of multiple exponential distributions can resemble a power-law WTD.

In principle, SEPs are one of representatives of solar eruptions. Therefore, the study of WTDs of SEP events, from another standpoint, may provide important clues for understanding the relationship between SEPs and solar eruptions. Another aspect motives this study is to generalize the model of Wheatland (2000) and Aschwanden $\&$ McTiernan (2010). We find that the generalized model fits well with the WTDs of SEP events, and it might also be applicable for the WTDs of other SOC systems.

\section{Observations}

The SEPs mainly consist of electrons and protons, with a small amount of high-Z ions. The solar electron events (SEEs) are selected based on the observations of the $WIND$ spacecraft, which was launched on 1994 November 1 and operates to the present time orbiting around Sun-Earth Lagrange 1 (L1) point. The three-dimensional Plasma and Energetic Particle instrument (3DP; Lin et al. 1995) observes electrons with electron electrostatic analyzers (EESAs) in energy range $\sim$3 eV - 30 KeV and with solid-state telescopes (SSTs) in $\sim$ 25 - 400 keV. Our survey of the data began in 1995 January and continued through 2012 December, this leads to the identification of 1594 SEEs with energies between $\sim$0.1 - 310 keV. Note that the orbit of $WIND$ had passed through the Earth's magnetosphere 73 times during the statistical period, it is necessary to remove these artificial waiting times. A statistical study of SEEs during solar cycle 23 has been done by Wang et al. (2012), the candidates studied in present work are the extension of the previous study.

The solar proton events (SPEs) represent more energetic solar eruptions and determine important properties of space weather. A SPE used to be defined as a flux of $\geq$10 MeV protons greater than 1 pfu (particle flux unit, 1 pfu = 1 $particle\ cm^{-2}\ s^{-1}\ sr^{-1}$). For some practical reasons (see discussion in Miroshnichenko 2003), the NOAA Space Environment Service Center (SESC) suggested to use an intensity of 10 pfu for $\geq$10 MeV protons as a reliable signature of SPEs. In the context of representatives of more energetic solar eruptions, we use the NOAA SESC criterion for a SPE. According to the survey of NOAA SESC, a total number of 252 SPEs were recorded by the Geostationary Operational Environment Satellite (GOES) from 1976 January to 2013 December ($http$://$www.swpc.noaa.gov/ftpdir/indices/SPE.txt$). The type II radio bursts are observed by 10 radio observatories around the world that cover all time zones ($ftp$://$ftp.ngdc.noaa.gov/STP/space$-$weather/solar$-$data/solar$-$features/solar$-$radio/radio$-$bursts/$). The survey of data began in 1995 January through to 2010 December, this leads to the identification of 1076 events.

\section{Methods and results}

\subsection{Non-stationary Poisson process}

Theoretically, a SOC event occurs independently and randomly, that follows a standard Poisson process and predicts an exponential WTD:
\begin{equation}\label{1}
P(\Delta t) = \lambda e^{-\lambda \Delta t},
\end{equation}
where $\Delta t$ describes the waiting time, and $\lambda$ the mean event occurrence rate. If the event rate varies time $\lambda(t)$, following the derivation of Wheatland (2000) and Aschwanden $\&$ McTiernan (2010), the time-dependent or non-stationary Poisson process will give the WTD as
\begin{equation}\label{2}
P(t, \Delta t) = \frac{\int^{T}_{0}\lambda(t)^{2}e^{-\lambda(t)\Delta t}dt}{\int_{0}^{T}\lambda(t)dt},
\end{equation}
where the observation time interval is [0, T], if t $>$ T, then $\lambda(t) = 0$.

Defining $f(\lambda)d\lambda = dt/T$, where $f(\lambda)$ is the fraction of time that the event rate is in the range ($\lambda$, $\lambda + d\lambda$), in other words the time distribution of event rate, a more tractable expression can be given:
\begin{equation}\label{3}
P(\Delta t) = \frac{\int_{0}^{\infty}f(\lambda)\lambda^{2}e^{-\lambda \Delta t}d\lambda}{\int_{0}^{\infty}\lambda f(\lambda)d\lambda}.
\end{equation}
The mean event rate is $\overline{\lambda}=\int_{0}^{\infty}\lambda f(\lambda)d\lambda$ during the observation time interval [0, T].

Comparing the solar-cycle distribution of sunspots, the SEEs display two distinct features as shown in Figure 1: (1) the maximum of the SEEs is delayed $\sim$1.5 years, and (2) the high intermittency (the short-term fluctuations are stronger than that of the sunspot number) is clearly recognized. The high intermittency reflects SEEs occur in a form of clustering, similar to the phenomenon of flare distribution (Aschwanden $\&$ McTiernan 2010). According to Wheatland (2000), the clusterization suggests that the time distribution of event rate $f(\lambda)$ may follow an exponential function. Here we apply a form of $f(\lambda)$ as
\begin{equation}\label{4}
f(\lambda) = A \lambda^{-\alpha}exp(-\beta \lambda).
\end{equation}
This is a generalized form of $f(\lambda)$ that may fits well with the real observations. This form includes the one of Wheatland (2000) who suggests an exponential form of $f(\lambda)$ where $\alpha = 0$. This form also includes the Case (4) and (5) of Aschwanden $\&$ McTiernan (2010) where $\alpha = 0$ and $\alpha = 1$, respectively. Taking Equation 4 into 3, we derive the analytical expression of WTD:
\begin{equation}\label{3}
P(\Delta t) = A \frac{\Gamma(3 - \alpha)}{\overline{\lambda}}(\beta + \Delta t)^{-(3 - \alpha)},
\end{equation}
where $\beta = (\frac{\Gamma(2 - \alpha)}{\overline{\lambda}})^{\frac{1}{2 - \alpha}}$, and $0 \leq \alpha < 2$. The constant $\Gamma$ corresponds to  the so called gamma function.

\subsection{WTDs of SEEs and SPEs}

Figure 2 shows the WTD of SEEs. We fit the WTD in two forms: a broken power law (left panel) and a non-stationary Poisson distribution with Equation 5 (right panel). The break of the power law occurs at $\sim$70 $< \Delta t <$ $\sim$100 hours. The power-law index is $\gamma_{1} =$ 0.99 for waiting times $<$70 hours and $\gamma_{2} =$ 1.92 for waiting times $>$100 hours. An interesting phenomenon is that the break time is consistent with the periodical crossings of the stream interfaces (SIs) or the corotating interaction regions (CIRs) where a fast solar wind stream overtake a leading slow one. A pair of shock waves (forward and reverse shocks) may form at the edges of CIRs (Gosling $\&$ Pizzo 1999), where charged particles can be energized (Reames 1999). This may explain the ``bump" around the break time and the origin of the broken power law WTD of SEEs.

The best fit of WTD with Equation 5 gives $\alpha = 1.46$ (right panel of Figure 2). Note that the waiting times over 1000 hours are ignored in order to minimize the fitting errors. Figure 3 shows the fitting of WTD for waiting times $<$70 hours (red dashed line), it leads to $\alpha_{1} = 1.46$. Then we fit the WTD for waiting times $>$100 hours (blue dashed line) in a form of $P(\Delta t, \alpha_{1}=1.46) + P(\Delta t, \alpha_{2})$, it leads to $\alpha_{2} = 0.39$. It is clear that the same value of index $1.46$ is obtained for all waiting times and waiting times $<$70 hours. This indicates that the deviation of WTD for waiting times $>$100 hours may arise from the same mechanism of the broken power law WTD caused by the modulation of CIRs.

A comparison of WTDs of SPEs and type II radio bursts is shown in Figure 4. Both WTDs are fitted in two forms: a power law for long waiting times and a non-stationary Poisson distribution with Equation 5. Very similar profiles of WTDs are recognized between SPEs and type II radio bursts. The power law index $\gamma$ is 1.82 for SPEs and 1.83 for type II radio bursts. The non-stationary Poisson distribution index $\alpha$ is 0.87 for SPEs and 0.82 for type II radio bursts. This confirms the close relationship between the proton acceleration and the shock wave that is responsible for the formation of the type II radio bursts.

\section{Summary and discussion}

Waiting time statistics of SEP events are investigated in present study. The WTDs of SEEs and SPEs are consistent with a non-stationary Poisson process, which was proposed to explain the WTD of solar flares (Wheatland 2000; Aschwanden $\&$ McTiernan 2010). A generalized non-stationary Poisson distribution is derived to interpret the WTDs of SEP events. Our conclusions are as following.

(1) The solar-cycle distribution of SEEs is featured as clusterization. If the event rate $\lambda = 1/\Delta t$ varies as the time distribution of event rate $f(\lambda) = A \lambda^{-\alpha} exp(-\beta \lambda)$, the non-stationary Poisson process gives the WTD in a form of Equation 5. That predicts a power-law tail $\sim \Delta t^{\alpha - 3}$, where $0 \leq \alpha < 2$.

(2) The WTD of SEEs show a broken power-law profile. The power-law index is $\gamma_{1} =$ 0.99 for waiting times $<$70 hours and $\gamma_{2} =$ 1.92 for waiting times $>$100 hours. This might be due to the modulation of CIRs.

(3) The WTD of SPEs can be well fitted with Equation 5. Similar WTDs of SPEs and type II radio bursts are recognized, that indicates a close relationship between proton acceleration and the shock waves responsible for the formation of type II radio bursts.

To be noticed that the power-law index was derived to be 2.16 for GOES soft X-ray flares(Wheatland 2000) and was $\sim$2.0 for RHESSI hard X-ray flares (Aschwanden $\&$ McTiernan 2010). In present study, the power-law index of WTD of SEEs is 0.99 for short waiting times, that is much harder than the WTD of flares. The reason probably arises from the fact that SEEs are not only flare-related, but many of them appear to be associated with narrow CMEs, coronal jets, or energy releases in high coronal sites (Kahler et al. 2001; Klein et al. 2001; Pick et al. 2006; Li et al. 2011; Wang et al. 2012). The power-law index of WTD of SEEs is 1.92 for long waiting times, taking into account the possible modulation of CIRs, that is comparable to the WTDs of flares and SPEs. This indicates the relationship between the solar eruption and the production of SEPs.

\begin{acknowledgements}
We are grateful to the $GOES$ and $WIND$ teams for providing observational data in this study. This work is supported by the project 985 of Nanjing University and Advanced Discipline Construction Project of Jiangsu Province and NKBRSF under grants 2014CB744203. C. Li would like to thank the Natural Science Foundation (BK2012299) of Jiangsu province and NSFC under grants 11303017. L. Wang was supported in part by NSFC under contract 41274172.
\end{acknowledgements}

\begin{figure}
   \centering
   \includegraphics[width=8cm]{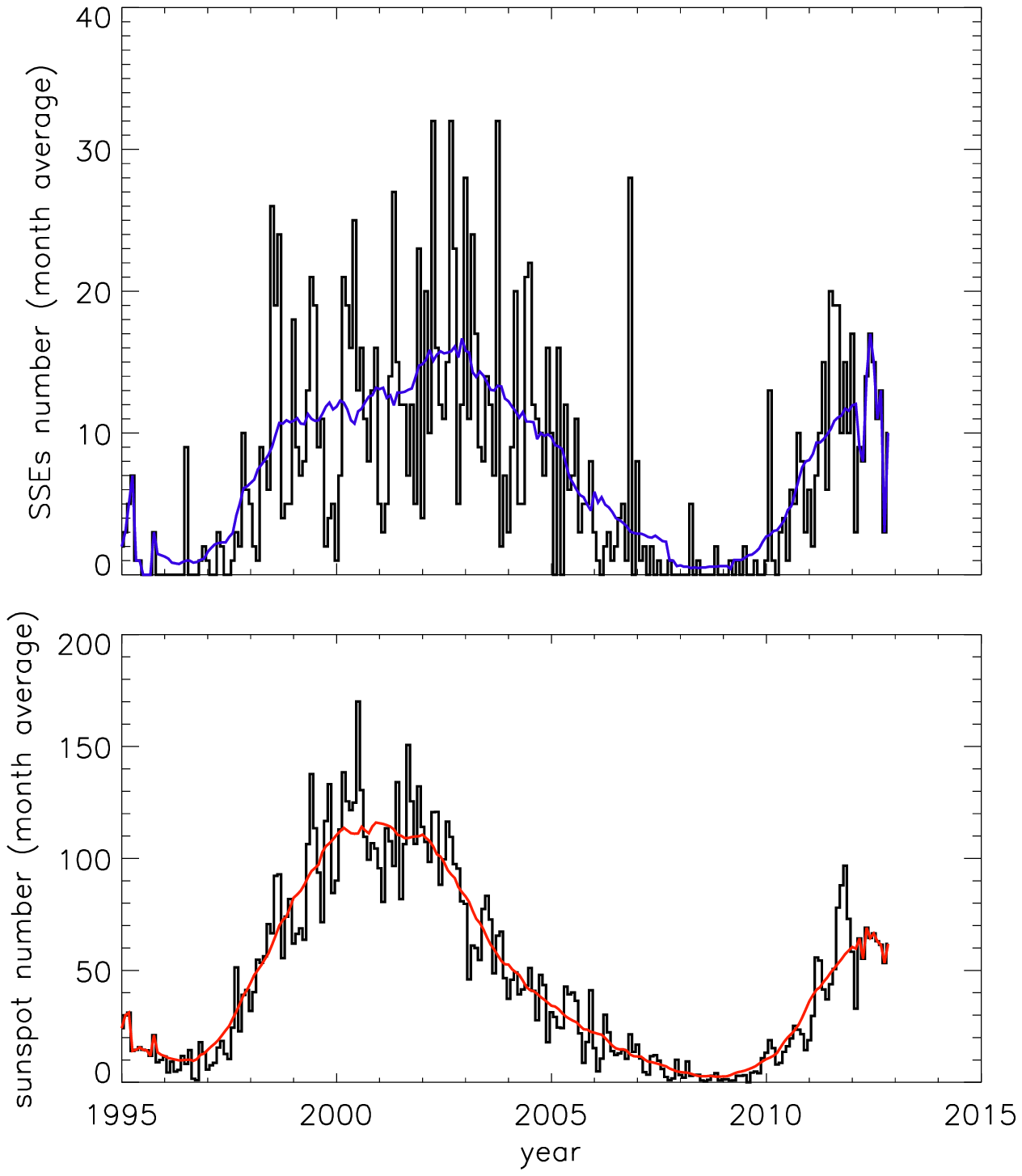}
   \caption{Distributions of SEEs and sunspots from 1995 January to 2012 December.}
   \label{fig1}
\end{figure}

\begin{figure*}
     \centering
     \vspace{0.0\textwidth}    % Shift back to the panel bottom
     \centerline{
               \hspace*{0.00\textwidth}
               \includegraphics[width=0.4\textwidth]{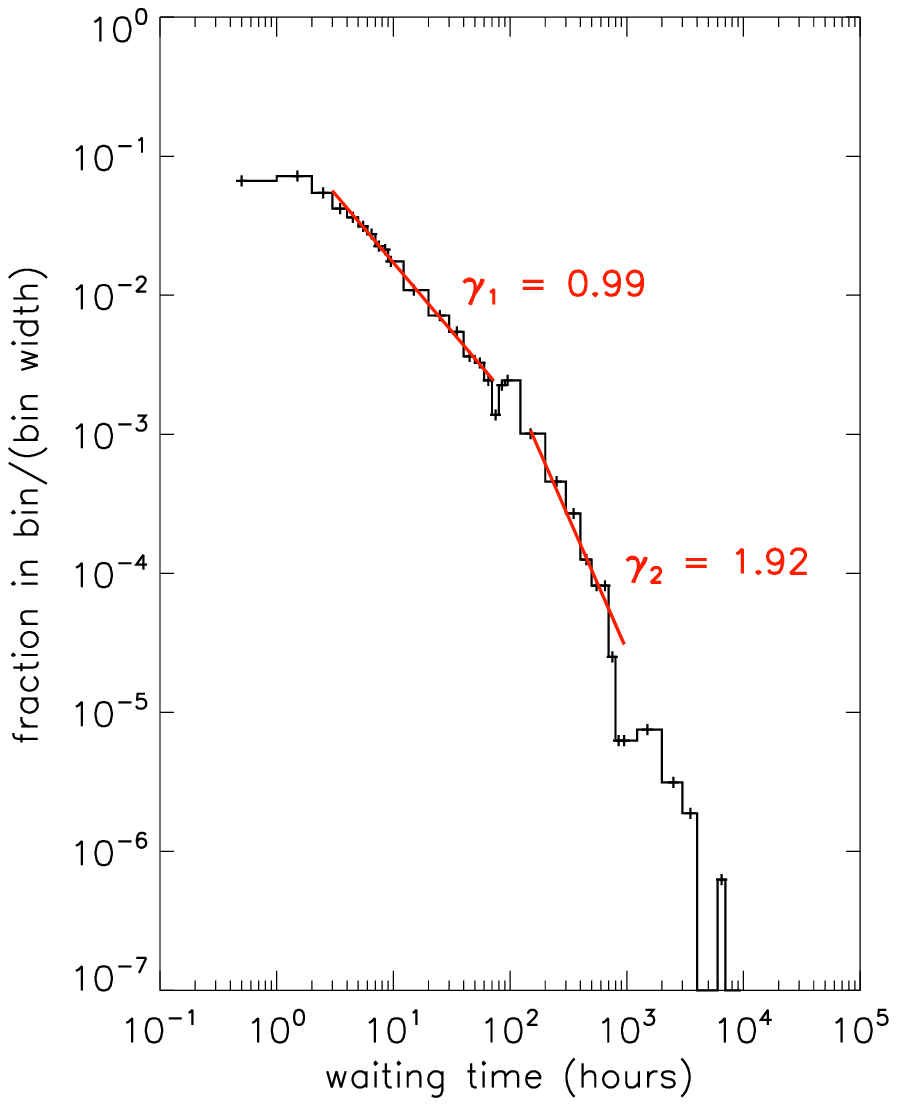}
               \includegraphics[width=0.4\textwidth]{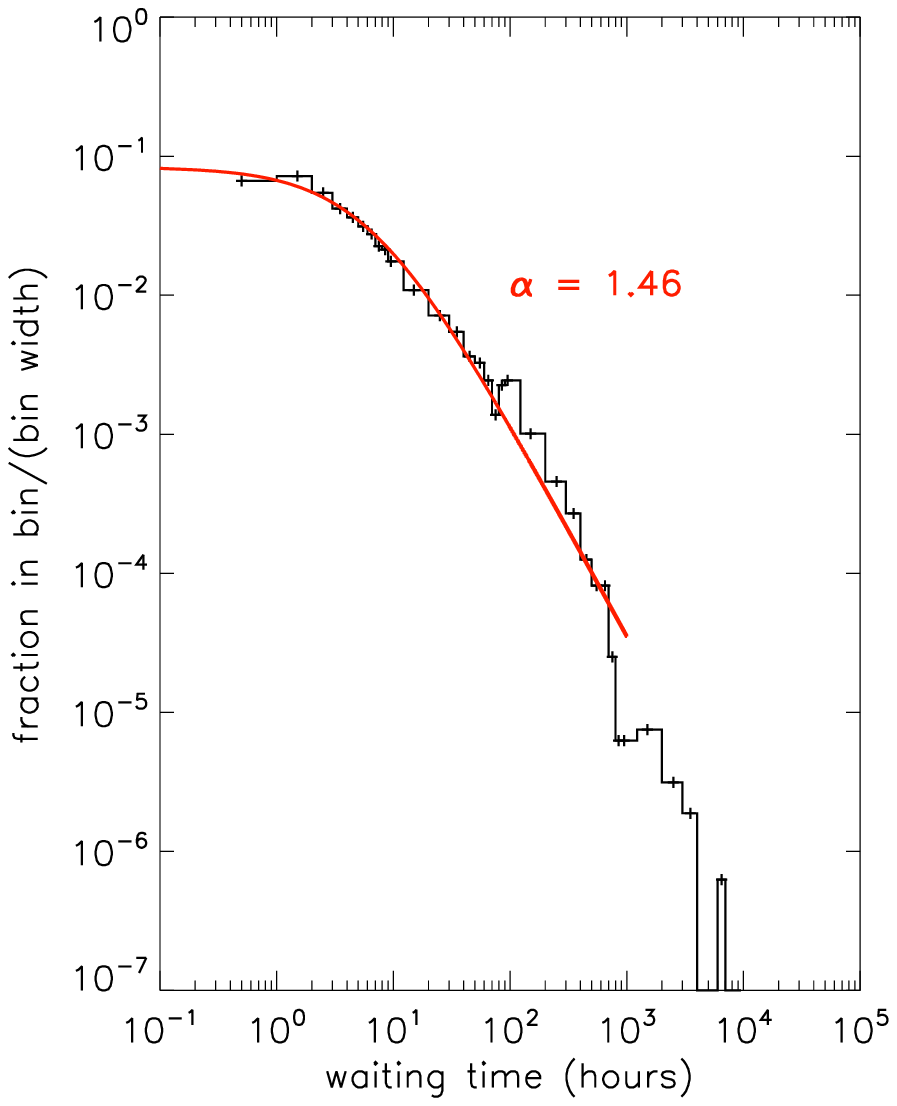}
               }
     \vspace{0.0\textwidth}   % Shift close to the panel top
\caption{WTD of SEEs fitted with a broken power law (left panel) and a non-stationary Poisson distribution with Equation 5 (right panel).}\label{fig2}
\end{figure*}

\begin{figure}
   \centering
   \includegraphics[width=8cm]{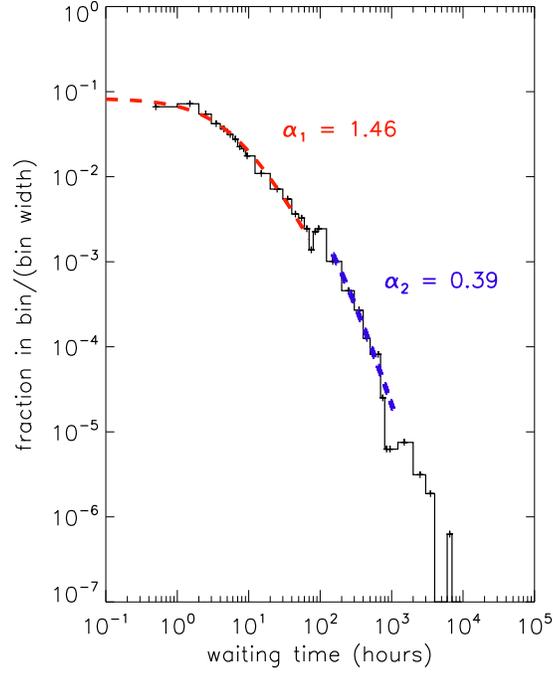}
   \caption{WTD of SEEs fitted with Equation 5 for waiting times $<$70 hours and $>$100 hours, respectively.}
   \label{fig3}
\end{figure}

\begin{figure*}
     \centering
     \vspace{0.0\textwidth}    % Shift back to the panel bottom
     \centerline{
               \hspace*{0.00\textwidth}
               \includegraphics[width=0.4\textwidth]{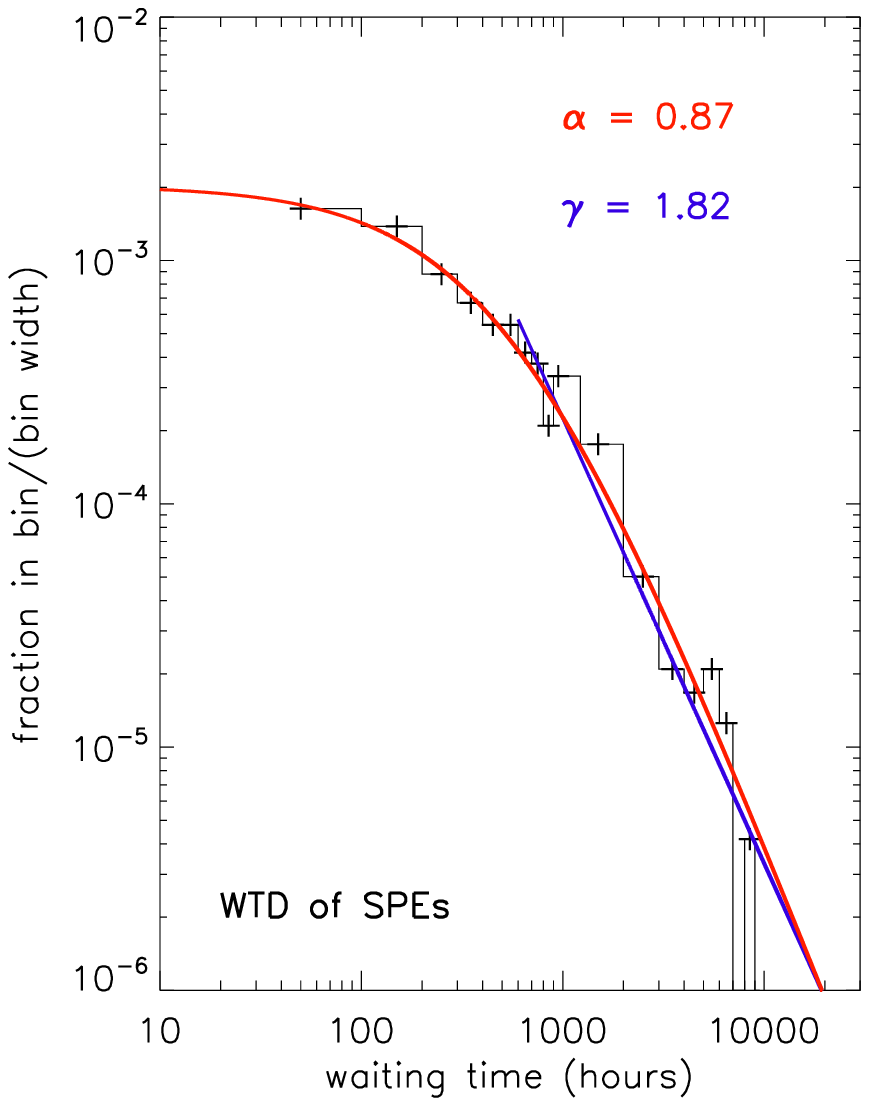}
               \includegraphics[width=0.4\textwidth]{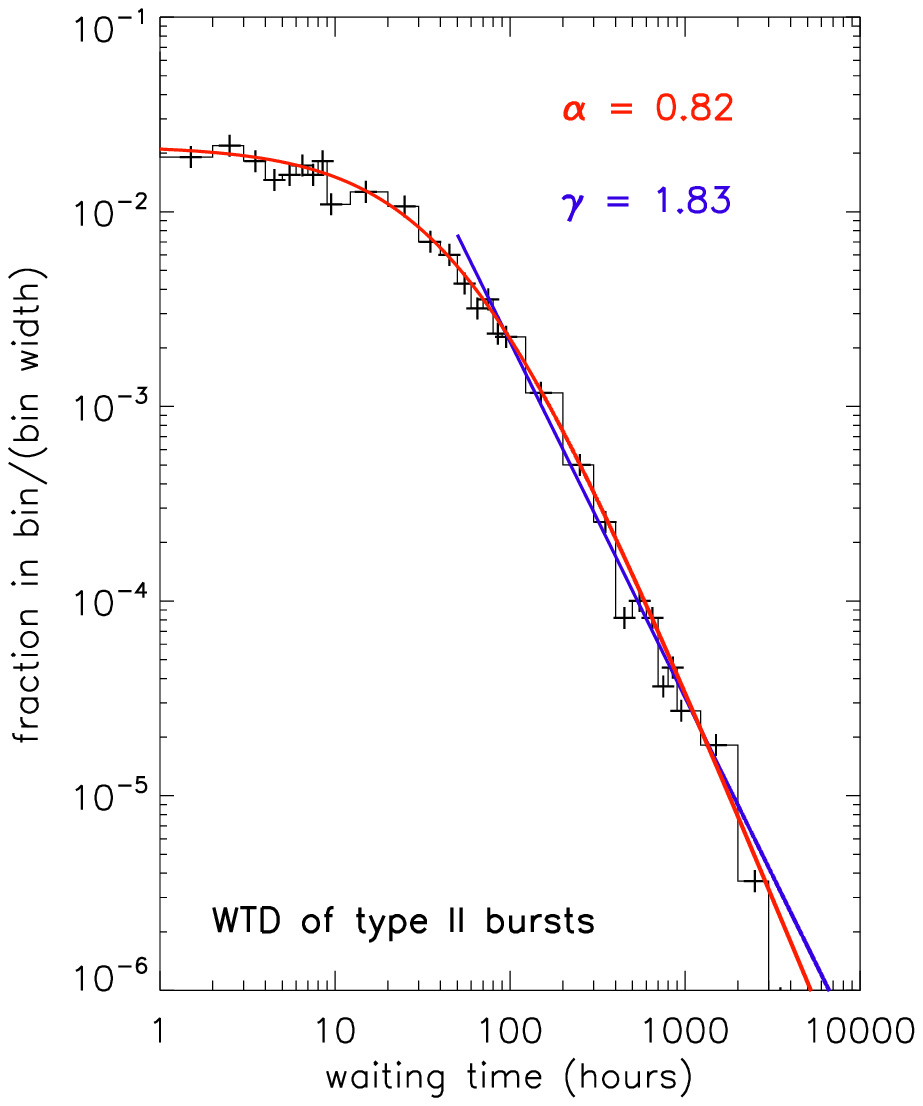}
               }
     \vspace{0.0\textwidth}   % Shift close to the panel top
\caption{WTDs of SPEs (left panel) and type II radio bursts (right panel). The blue line indicates power-law fitting and red line fitting with Equation 5.}\label{fig4}
\end{figure*}

\end{document}